\magnification=1200
\def\qed{\unskip\kern 6pt\penalty 500\raise -2pt\hbox
{\vrule\vbox to 10pt{\hrule width 4pt\vfill\hrule}\vrule}}
\centerline{GRACE-LIKE POLYNOMIALS.}
\bigskip\bigskip
\centerline{by David Ruelle\footnote{*}{IHES.  91440 Bures sur Yvette,
France. $<$ruelle@ihes.fr$>$}.}
\bigskip\bigskip\bigskip\bigskip\noindent
	{\leftskip=2cm\rightskip=2cm\sl Abstract.  Results of somewhat mysterious nature are known on the location of zeros of certain polynomials associated with statistical mechanics (Lee-Yang circle theorem) and also with graph counting.  In an attempt at clarifying the situation we introduce and discuss here a natural class of polynomials.  Let $P(z_1,\ldots,z_m,w_1,\ldots,w_n)$ be separately of degree 1 in each of its $m+n$ arguments.  We say that $P$ is a Grace-like polynomial if $P(z_1,\ldots,w_n)\ne0$ whenever there is a circle in ${\bf C}$ separating $z_1,\ldots,z_m$ from $w_1,\ldots,w_n$.  A number of properties and characterizations of these polynomials are obtained.\par}
\bigskip\bigskip\bigskip\bigskip\noindent
\vfill\eject
\bigskip\bigskip
	{\sl I had the luck to meet Steve Smale early in my scientific career, and I have read his 1967 article in the Bulletin of the AMS more times than any other scientific paper.  It took me a while to realize that Steve had worked successively on a variety of subjects, of which ``differentiable dynamical systems'' was only one.  Progressively also I came to appreciate his independence of mind, expressed in such revolutionary notions as that the beaches of Copacabana are a good place to do mathematics.  Turning away from scientific nostalgy, I shall now discuss a problem which is not very close to Steve's work, but has relations to his interests in recent years: finding where zeros of polynomials are located in the complex plane.}
\medskip
	{\bf 0 Introduction.} 
\medskip
	One form of the Lee-Yang circle theorem [3] states that if $|a_{ij}|\le1$ for $i,j=1,\ldots,n$, and $a_{ij}=a^*_{ji}$, the polynomial
$$	\sum_{X\subset\{1,\ldots,n\}}z^{{\rm card}X}
	\prod_{i\in X}\prod_{j\notin X}a_{ij}      $$
has all its zeros on the unit circle $\{z:|z|=1\}$.
\medskip
	Let now $\Gamma$ be a finite graph.  We denote by $\Gamma'$ the set of dimer subgraphs $\gamma$ (at most one edge of $\gamma$ meets any vertex of $\Gamma$), and by $\Gamma''$ the set of unbranched subgraphs $\gamma$ (no more than two edges of $\gamma$ meet any vertex of $\Gamma$).  Writing $|\gamma|$ for the number of edges in $\gamma$, on proves that 
$$	\sum_{\gamma\in\Gamma'}z^{|\gamma|}      $$
has all its zeros on the negative real axis (Heilmann-Lieb [2]) and 
$$	\sum_{\gamma\in\Gamma''}z^{|\gamma|}      $$
has all its zeros in the left-hand half plane $\{z:{\rm Im}z<0\}$ (Ruelle [6]).
\medskip
	The above results can all be obtained in a uniform manner by studying the zeros of polynomials 
$$	P(z_1,\ldots,z_n)      $$
which are {\it multiaffine} (separately of degree $1$ in their $n$ variables), and then taking $z_1=\ldots=z_n=z$.  The multiaffine polynomials corresponding to the three examples above are obtained by multiplying factors for which the location of zeros is known and performing {\it Asano contractions}:
$$	Auv+Bu+Cv+D\qquad\rightarrow\qquad Az+D      $$
The {\it key lemma} (see [5]) is that if $K$, $L$ are closed subsets of ${\bf C}\backslash\{0\}$ and if 
$$	u\notin K, v\notin L\qquad\Rightarrow\qquad Auv+Bu+Cv+D\ne0      $$
then
$$	z\notin-K*L\qquad\Rightarrow\qquad Az+D\ne0      $$
where we have written $K*L=\{uv:u\in K,v\in L\}$.
\medskip
	To get started, let $P(z_1,\ldots,z_n)$ be a multiaffine {\it symmetric} polynomial.  If $W_1,\ldots,W_n$ are the roots of $P(z,\ldots,z)=0$, we have 
$$   P(z_1,\ldots,z_n)={\rm const.}\sum_\pi\prod_{j=1}^n(z_j-W_{\pi(j)})   $$
where the sum is over all permutations $\pi$ of $n$ objects.  {\it Grace's theorem} asserts that if $Z_1,\ldots,Z_n$ are separated from $W_1,\ldots,W_n$ by a circle of the Riemann sphere, then $P(Z_1\ldots,Z_n)\ne0$.  For example, if $a$ is real and $-1\le a\le1$, the roots of $z^2+2az+1$ are on the unit circle, and therefore
$$	uv+au+av+1      $$
cannot vanish when $|u|<1$, $|v|<1$; from this one can get the Lee-Yang theorem.
\medskip
	In view of the above, it is natural to consider multiaffine polynomials
$$	P(z_1,\ldots,z_m,w_1,\ldots,w_n)      $$
which cannot vanish when $z_1,\ldots,z_m$ are separated from $w_1,\ldots,w_n$ by a circle.  We call these polynomials Grace-like, and the purpose of this note is to study and characterize them.
\bigskip
\centerline{\bf I. General theory.}
\bigskip
	We say that a complex polynomial $P(z_1,z_2,\ldots)$ in the variables $z_1$, $z_2,\ldots$ is a Multi-Affine Polynomial ({\it MA-nomial} for short) if it is separately of degree 1 in $z_1$, $z_2,\ldots$.  We say that a circle $\Gamma\subset{\bf C}$ {\it separates} the sets $A'$, $A''\subset{\bf C}$ if ${\bf C}\backslash\Gamma=C'\cup C''$, where $C'$, $C''$ are open, $C'\cap C''=\emptyset$ and $A'\subset C'$, $A''\subset C''$.  We say that the MA-nomial $P(z_1,\ldots,z_m,w_1,\ldots,w_n)$ is Grace-like (or a G-nomial for short) if it satisfies the following condition
\medskip
	(G) {\sl Whenever there is a circle $\Gamma$ separating $\{Z_1,\ldots,Z_m\}$, $\{W_1,\ldots,W_n\}$, then}
$$	P(Z_1,\ldots,W_n)\ne0      $$
	[Note that we call circle either a straight line $\Gamma\subset{\bf R}$ or a {\it proper circle} $\Gamma=\{z\in{\bf C}:|z-a|=R\}$ with $a\in{\bf C}$, $0<R<\infty$].
\medskip
	{\bf 1 Lemma} (homogeneity).  
\medskip
	{\sl The G-nomial $P$ is homogeneous of degree $k\le\min(m,n)$.}
\medskip
	If there is a circle $\Gamma$ separating $\{z_1,\ldots,z_m\}$, $\{w_1,\ldots,w_n\}$, then the polynomial $\lambda\mapsto P(\lambda z_1,\ldots,\lambda w_n)$ does not vanish when $\lambda\ne0$, hence it is of the form $C\lambda^k$, where $C=P(z_1,\ldots,w_n)$.  Thus
$$   P(\lambda z_1,\ldots,\lambda w_n)=\lambda^kP(z_1,\ldots,w_n)   $$
on an open set of ${\bf C}^{m+n}$, hence identically, {\it i.e.}, $P$ is homogeneous of degree $k$.
\medskip
	Assuming $k>n$, each monomial in $P$ would have a factor $z_i$, hence 
$$	P(0,\ldots,0,1,\ldots,1)=0      $$ 
in contradiction with the fact that $\{0,\ldots,0\}$, $\{1,\ldots,1\}$ are separated by a circle.  Thus $k\le n$, and similarly $k\le m$.\qed
\medskip
	{\bf 2 Lemma} (degree).  
\medskip
	{\sl If all the variables $z_1,\ldots,w_n$ effectively occur in the G-nomial $P$, then $m=n$ and $P$ has degree $k=n$.}
\medskip
	By assumption
$$   (\prod_{i=1}^mz_i)(\prod_{j=1}^nw_j)P(z_1^{-1},\ldots,w_n^{-1})   $$
is a homogeneous MA-nomial $\tilde P(z_1,\ldots,w_n)$ of degree $m+n-k$.  If $Z_1,\ldots,W_n$ are all $\ne0$ and $\{Z_1,\ldots,Z_m\}$, $\{W_1,\ldots,W_n\}$ are separated by a circle $\Gamma$, we may assume that $\Gamma$ does not pass through $0$.  Then $\{Z_1^{-1},\ldots,Z_m^{-1}\}$, $\{W_1^{-1},\ldots,W_n^{-1}\}$ are separated by $\Gamma^{-1}$, hence $\tilde P(Z_1,\ldots,W_n)\ne0$.  Let ${\cal V}$ be the variety of zeros of $\tilde P$ and 
$$	{\cal Z}_i=\{(z_1,\ldots,w_n):z_i=0\}\qquad,
	\qquad{\cal W}_j=\{(z_1,\ldots,w_n):w_j=0\}      $$
Then
$$	{\cal V}\subset({\cal V}\backslash\cup_{i,j}({\cal Z}_i\cup{\cal W}_j))
	\cup\cup_{i,j}({\cal Z}_i\cup{\cal W}_j)      $$
Since all the variables $z_1,\ldots,w_n$ effectively occur in $P(z_1,\ldots,w_n)$, none of the hyperplanes ${\cal Z}_i$, ${\cal W}_j$ is contained in ${\cal V}$, and therefore
$$	{\cal V}\subset
\hbox{closure}({\cal V}\backslash\cup_{i,j}({\cal Z}_i\cup{\cal W}_j))   $$
We have seen that the points $(Z_1,\ldots,W_n)$ in ${\cal V}\backslash\cup_{i,j}({\cal Z}_i\cup{\cal W}_j)$ are such that $\{Z_1,\ldots,Z_m\}$, $\{W_1,\ldots,W_m\}$ cannot be separated by a circle $\Gamma$, and the same applies to their limits.  Therefore $\tilde P$ again satisfies (G).  Applying Lemma 1 to $P$ and $\tilde P$ we see that $k\le\min(m,n)$, $m+n-k\le\min(m,n)$.  Therefore $m+n\le2\min(m,n)$, thus $m=n$, and also $k=n$.\qed
\medskip
	{\bf 3 Proposition} (reduced G-nomials).
\medskip
	{\sl If $P(z_1,\ldots,z_m,w_1,\ldots,w_n)$ is a G-nomial, then $P$ depends effectively on a subset of variables which may be relabelled $z_1,\ldots,z_k,w_1,\ldots,w_k$ so that 
$$	P(z_1,\ldots,z_m,w_1,\ldots,w_n)
	=\alpha R(z_1,\ldots,z_k,w_1,\ldots,w_k)      $$
where $\alpha\ne0$, the G-nomial $R$ is homogeneous of degree $k$, and the coefficient of $z_1\cdots z_k$ in $R$ is 1.}
\medskip
	This follows directly from Lemma 2.\qed
\medskip
	We call a G-nomial $R$ as above a {\it reduced} G-nomial.
\medskip
	{\bf 4 Lemma} (translation invariance).  
\medskip
	{\sl If $P(z_1,\ldots,w_n)$ is a G-nomial, then
$$	P(z_1+s,\ldots,w_n+s)=P(z_1,\ldots,w_n)      $$
{\it i.e.,} $P$ is translation invariant.}
\medskip
	If  there is a circle $\Gamma$ separating $\{z_1,\ldots,z_m\}$, $\{w_1,\ldots,w_n\}$, then the polynomial
$$	p(s)=P(z_1+s,\ldots,w_n+s)      $$
satisfies $p(s)\ne0$ for all $s\in{\bf C}$.  This implies that $p(s)$ is constant, or $dp/ds=0$, for $(z_1,\ldots,w_n)$ in a nonempty open subset of ${\bf C}^{2n}$.  Therefore $dp/ds=0$ identically, and $p$ depends only on $(z_1,\ldots,w_n)$.  From this the lemma follows.\qed
\medskip
	{\bf 5 Proposition} (properties of reduced G-nomials).
\medskip
	{\sl If $P(z_1,\ldots,w_n)$ is a reduced G-nomial, the following properties hold:
\medskip
	{\rm (reduced form)} there are constants $C_\pi$ such that $P$ has the reduced form
$$	P(z_1,\ldots,w_n)=\sum_\pi C_\pi\prod_{j=1}^n(z_j-w_{\pi(j)})      $$
where the sum is over all permutations $\pi$ of $(1,\ldots,n)$
\medskip
	{\rm (conformal invariance)} if $ad-bc\ne0$, then
$$	P({az_1+b\over cz_1+d},\ldots,{aw_n+b\over cw_n+d})
	=P(z_1,\ldots,w_n)\prod_{j=1}^n{ad-bc\over(cz_j+d)(cw_j+d)}      $$
in particular we have  the identity
$$	(\prod_{i=1}^kz_i)(\prod_{j=1}^kw_j)R(z_1^{-1},\ldots,w_k^{-1})
	=(-1)^kR(z_1,\ldots,w_k)      $$
\indent
	{\rm (roots)} the polynomial
$$	\hat P(z)=P(z,\ldots,z,w_1,\ldots,w_n)      $$
is equal to $\prod_{k=1}^n(z-w_k)$, so that its roots are the $w_k$ (repeated according to multiplicity).}
\medskip
	Using Proposition 3 and Lemma 4, the above properties follow from Proposition A2 and Corollary A3 in Appendix A.\qed
\medskip
	{\bf 6 Proposition} (compactness).
\medskip
	{\sl The space of MA-nomials in $2n$ variables which are homogeneous of degree $n$ may be identified with ${\bf C}^{({2n\atop n})}$.  The set $G_n$ of reduced G-nomials of degree $n$ is then a compact subset of ${\bf C}^{({2n\atop n})}$.  We shall see later (Corollary 15) that $G_n$ is also contractible.}
\medskip
	Let $P_k\in G_n$ and $P_k\to P_\infty$.  In particular $P_\infty$ is homogeneous of degree $n$, and the monomial $z_1\cdots z_n$ occurs with coefficient 1.  Suppose now that
$$	P_\infty(Z_1,\ldots,Z_m,W_1,\ldots,W_n)=0      $$
with $\{Z_1,\ldots,Z_m\}$, $\{W_1,\ldots,W_n\}$ separated by a circle $\Gamma$.  One can then choose discs $D_1$, \dots, $D_{2n}$ containing $\{Z_1,\ldots,W_n\}$ and not intersecting $\Gamma$.  Lemma A1 in Appendix A would then imply that $P_\infty$ vanishes identically in contradiction with the fact that $P_\infty$ contains the term $z_1\cdots z_n$.  Therefore $P_\infty\in G_n$, {\it i.e.}, $G_n$ is closed.
\medskip
	Suppose now that $G_n$ were unbounded.  There would then be $P_k$ such that the largest coefficient (in modulus) $c_k$ in $P_k$ tends to $\infty$.  Going to a subsequence we may assume that
$$	c_k^{-1} P_k\to P_\infty      $$
where $P_\infty$ is a homogeneous MA-nomial of degree $n$, and does not vanish identically.  The same argument as above shows that $P_\infty$ is a G-nomial, hence (by Proposition 3) the coefficient $\alpha$ of $z_1\cdots z_n$ does not vanish, but since $\alpha=\lim c_k^{-1}$, $c_k$ cannot tend to infinity as supposed.  $G_n$ is thus bounded, hence compact.\qed
\medskip
	{\bf 7 Proposition} (the cases $n=1$, $2$).
\medskip
	{\sl The reduced G-nomials with $n=1,2$ are as follows:
\medskip\noindent
	For $n=1$: $P=z_1-w_1$.
\medskip\noindent
	For $n=2$: $P=(1-\theta)(z_1-w_1)(z_2-w_2)+\theta(z_1-w_2)(z_2-w_1)$ with real $\theta\in[0,1]$.}
\medskip
	We use Proposition 5 to write $P$ in reduced form.
\medskip\noindent
	In the case $n=1$, we have $P=C(z_1-w_1)$, and $C=1$ by normalization. 
\medskip\noindent
	In the case $n=2$, we have
$$	P=C'(z_1-w_1)(z_2-w_2)+C''(z_1-w_2)(z_2-w_1)      $$
In view of (G), $C'$, $C''$ are not both $0$.  Assume $C'\ne0$, then (G) says that 
$$	{z_1-w_1\over z_1-w_2}:{z_2-w_1\over z_2-w_2}
	+{C''\over C'}\ne0\eqno{(1)}      $$
when $\{z_1,z_2\}$, $\{w_1,w_2\}$ are separated by a circle.  If $C''/C'$ were not real, we could find $z_1,z_2,w_1,w_2$ such that 
$$	{z_1-w_1\over z_1-w_2}:{z_2-w_1\over z_2-w_2}
	=-{C''\over C'}\eqno{(2)}      $$
but the fact that the cross-ratio in the left hand side of (2) is not real means that $z_1,z_2,w_1,w_2$ are not on the same circle, and this implies that there is a circle separating $\{z_1,z_2\}$, $\{w_1,w_2\}$.  Therefore (1) and (2) both hold, which is impossible.  We must therefore assume $C''/C'$ real, and it suffices to check (1) for $z_1,z_2,w_1,w_2$ on a circle.  The condition that $\{z_1,z_2\}$, $\{w_1,w_2\}$ are separated by a circle is now equivalent to the cross-ratio being $>0$, and therefore (G) is equivalent to $C''/C'\ge0$.  If we assume $C''\ne0$, the argument is similar and gives $C'/C''\ge0$.  The normalization condition yields then $C'=1-\theta$, $C''=\theta$ with $\theta\in[0,1]$\qed
\medskip
	{\bf 8 Proposition} (determinants).
\medskip
	{\sl Let $\Delta_z$ be the diagonal $n\times n$ matrix where the $j$-th diagonal element is $z_j$ and similarly for $\Delta_w$.  Also let $U$ be a unitary $n\times n$ matrix ($U\Delta_wU^{-1}$ is thus an arbitrary normal matrix with eigenvalues $w_1,\ldots,w_n$).  Then
$$  P(z_1,\ldots,z_n,w_1,\ldots,w_n)=\det(\Delta_z-U\Delta_wU^{-1})  $$
is a reduced G-nomial.  We may assume that $\det U=1$ and write}
$$	\det(\Delta_z-U\Delta_wU^{-1})=\det((U_{ij}(z_i-w_j)))      $$
\medskip
	Let $\{z_1,\ldots,z_n\}$, $\{w_1,\ldots,w_n\}$ be separated by a circle $\Gamma$.  We may assume that $\Gamma$ is a proper circle.  Suppose first that the $z_j$ are inside the circle $\Gamma$ and the $w_j$ outside.  We want to prove that $\det(\Delta_z-U\Delta_wU^{-1})\ne0$.  By translation we may assume that $\Gamma$ is centered at the origin, say $\Gamma=\{z:|z|=R\}$; then, by assumption, using the operator norm, 
$$	||\Delta_z||<R\qquad,\qquad||\Delta_w^{-1}||<R^{-1}      $$
Therefore
$$	||\Delta_z(U\Delta_wU^{-1})^{-1}||<1      $$
so that
$$	\det(\Delta_z-U\Delta_wU^{-1})
=\det(-U\Delta_wU^{-1})\det(1-\Delta_z(U\Delta_wU^{-1})^{-1})\ne0      $$
as announced.  The case where the $w_j$ are inside $\Gamma$ and the $z_j$ outside is similar (consider $\det(\Delta_w-U^{-1}\Delta_zU)$.\qed
\medskip
	{\bf 9 Proposition} (Grace's theorem).
\medskip
	{\sl The polynomial
$$	P_\Sigma(z_1,\ldots,z_n,w_1,\ldots,w_n)
	={1\over n!}\sum_\pi\prod_{j=1}^n(z_j-w_{\pi(j)})\eqno{(3)}      $$
where the sum is over all permutations of $(1,\ldots,n)$ is a reduced G-nomial.}
\medskip
	See Polya and Szeg\"o [4] Exercise V 145.\qed
\medskip
	This result will also follow from our proof of Corollary 15 below.
\medskip
	{\bf 10 Proposition} (permanence properties).
\medskip
	(Permutations)  {\sl If $P(z_1,\ldots,z_n,w_1,\ldots,w_n)$ is a reduced G-nomial, permutation of the $z_i$, or the $w_j$, or interchange of $(z_1,\ldots,z_n)$ and $(w_1,\ldots,w_n)$ and multiplication by $(-1)^n$ produces again a reduced G-nomial.}
\medskip
	(Products)  {\sl If $P'(z'_1,\ldots,w'_{n'})$, $P''(z''_1,\ldots,w''_{n''})$ are reduced G-nomials, then their product $P'\otimes P''(z'_1,\ldots,z''_{n''},w'_1,\ldots,w''_{n''})$ is a reduced G-nomial.}
\medskip
	(Symmetrization)  {\sl Let $P(z_1,\ldots,z_n,w_1,\ldots,w_n)$ be a reduced G-nomial, and 
$$	P_S(z_1,\ldots,z_n,w_1,\ldots,w_n)      $$
be obtained by symmetrization with respect to a subset $S$ of the variables $z_1,\ldots,z_n$, then $P_S$ is again a reduced G-nomial.  Symmetrization with respect to all variables $z_1,\ldots,z_n$ produces the polynomial $P_\Sigma$ given by (3).}
\medskip
	The part of the proposition relative to permutations and products follows readily from the definitions.  To prove the symmetrization property we may relabel variables so that $S$ consists of $z_1,\ldots,z_s$.  We denote by $\hat P(z)$ the polynomial obtained by replacing $z_1,\ldots,z_s$ by $z$ in $P$ (the dependence on $z_{s+1},\ldots,w_n$ is not made explicit).  With this notation $P_S$ is the only MA-nomial symmetric with respect to $z_1,\ldots,z_n$ and such that $\hat P_S(z)=\hat P(z)$.  We may write
$$	\hat P(z)=\alpha(z-a_1)\cdots(z-a_s)\eqno{(4)}      $$
where $\alpha,a_1,\ldots,a_s$ may depend on $z_{s+1},\ldots,w_n$.  If $\Gamma$ is a circle separating the regions $C'$, $C''$, and $z_{s+1},\ldots,z_n\in C'$, $w_1,\ldots,w_n\in C''$, (G) implies that $\alpha\ne0$ and $a_1,\ldots,a_s\notin C'$.  Grace's theorem implies that $P_S$ does not vanish when $z_1,\ldots,z_s$ are separated by a circle from $a_1,\ldots,a_s$.  Therefore $P_S$ does not vanish when $z_1,\ldots,z_s\in C'$, hence $P_S$ is a G-nomial, which is easily seen to be reduced.  If $s=n$, (4) becomes
$$	\hat P(z)=(z-w_1)\cdots(z-w_n)      $$
in view of Proposition 5, hence symmetrisation of $P$ gives $P_\Sigma$.
\bigskip
\centerline{\bf II. Further results.}
\bigskip
	We define now G$_0$-nomials as a class of MA-nomials satisfying a new condition (G$_0$) weaker than (G).  It will turn out later that G$_0$-nomials and G-nomials are in fact the same (Proposition 12).  The new condition is 
\medskip
	(G$_0$) {\sl If there are two proper circles, or a proper circle and a straight line $\Gamma'$, $\Gamma''\subset{\bf C}$ such that $z_1,\ldots,z_m\in\Gamma'$,  $w_1,\ldots,w_n\in\Gamma''$, and $\Gamma'\cap\Gamma''=\emptyset$, then}
$$	P(z_1,\ldots,z_m,w_1,\ldots,w_n)\ne0      $$
\medskip
	Remember that a proper circle is of the form $\{z:|z-a|=R\}$, with $0<R<\infty$.  For the purposes of (G$_0$) we may allow $R=0$ (because a circle $\Gamma'$ or $\Gamma''$ reduced to a point $a'$ or $a''$ can be replaced by a small circle through $a'$ or $a''$).
\medskip
	{\bf 11 Lemma.}
\medskip
	{\sl Let $P(z_1,\ldots,w_n)$ be a G$_0$-nomial, and define
$$	\tilde P(z_1,\ldots,w_n)
=(\prod_{i=1}^m z_i)(\prod_{j=1}^n w_j)P(z_1^{-1},\ldots,w_n^{-1})\eqno{(5)}      $$
\indent
	(a) $P$ is translation invariant.
\medskip
	(b) If $P$ depends effectively on $z_1,\ldots,w_n$, then $\tilde P$ is translation invariant, and therefore a G$_0$-nomial.}
\medskip
	The polynomial $a\mapsto p(a)=P(z_1+a,\ldots,w_n+a)$ does not vanish, and is therefore constant if $z_1,\ldots,z_m\in\Gamma'$,  $w_1,\ldots,w_n\in\Gamma''$, and $\Gamma'\cap\Gamma''=\emptyset$.  But this means $dp/da=0$ under the same conditions, and therefore $dp/da$ vanishes identically.  This proves (a).
\medskip
	From the fact that $P$ depends effectively on $z_1,\ldots,w_n$, we obtain that none of the $m+n$ polynomials
$$	\tilde P(0,z_2-z_1,\ldots,w_n-z_1)      $$
$$	\ldots      $$
$$	\tilde P(z_1-w_n,\ldots,w_{n-1},0)      $$
vanishes identically.  The union ${\cal Z}$ of their zeros has thus a dense complement in ${\bf C}^{m+n}$.  Let now $\Gamma'$, $\Gamma''$ be disjoint proper circles in ${\bf C}$.  If $z_1,\ldots,z_m\in\Gamma'$,  $w_1,\ldots,w_n\in\Gamma''$, the polynomial
$$	a\mapsto\tilde p(a)=\tilde P(z_1+a,\ldots,w_n+a)      $$
can vanish only if $a\in\{-z_1,\ldots,-w_n\}$.  [This follows from (G$_0$) and the fact that $(a+\Gamma')^{-1}$, $(a+\Gamma'')^{-1}$ are disjoint and are proper circles or a proper circle and a straight line].  To summarize, $\tilde p(a)$ can vanish only if
$$	a\in\{-z_1,\ldots,-w_n\}\qquad{\rm and}
	\qquad (z_1,\ldots,w_n)\in {\cal Z}      $$
Since a polynomial vanishing on a nonempty open set of $\Gamma'^m\times\Gamma''^n$ must vanish identically on ${\bf C}^{m+n}$, we have 
$$	({\bf C}^{m+n}\backslash{\cal Z})\cap(\Gamma'^m\times\Gamma''^n)
	\ne\emptyset      $$
There is thus a nonempty open set $U\subset(\Gamma'^m\times\Gamma''^n)\backslash{\cal Z}$.  For $(z_1,\ldots,w_n)\in U$, $\tilde p(\cdot)$ never vanishes, and is thus constant, {\it i.e.}, $d\tilde p(a)/da=0$.  Therefore $d\tilde p(a)/da=0$ for all $(z_1,\ldots,w_n)\in{\bf C}^{m+n}$.   In conclusion, $\tilde P$ is translation invariant.  This implies immediately that $\tilde P$ is a G$_0$-nomial.\qed
\medskip
	{\bf 12 Proposition.}
\medskip
	{\sl If the MA-nomial $P(z_1,\ldots,z_m,w_1,\ldots,w_n)$ satisfies (G$_0$), it also satisfies (G).}
\medskip
	If the sets $\{z_1,\ldots,z_m\}$ and $\{w_1,\ldots,w_n\}$ are separated by a circle $\Gamma$, we can find two disjoint proper circles $\Gamma'$ and $\Gamma''$ close to $\Gamma$ and separating them.  By a transformation $\Phi:z\mapsto(z+a)^{-1}$, we may assume that $\Phi z_1,\ldots,\Phi z_m$ are {\it inside} of the circle $\Phi\Gamma'$, and $\Phi w_1,\ldots,\Phi w_n$ {\it inside} of the circle $\Phi\Gamma''$.
\medskip
	We may write 
$$	\Phi\Gamma'=\{z:|z-u|=r'\}\qquad,
	\qquad\Phi\Gamma''=\{w:|w-v|=r''\}      $$
The assumption that $P$ is a G$_0$-nomial and Lemma 11 imply that $\tilde P$ (defined by (5)) satisfies $\tilde P(z_1,\ldots,w_n)\ne0$ if
$$	z_1,\ldots,z_m\in\{z:|z-u|=\rho'\},\qquad,
	\qquad w_1,\ldots,w_n\in\{w:|w-v|=\rho''\}      $$
whenever $0\le\rho'\le r'$ and $0\le\rho''\le r''$.  Considered as a function of the $\xi_i=\log(z_i-u)$ and $\eta_j=\log(w_j-v)$, $\tilde P$ has no zero, and $1/\tilde P$ is thus analytic in a region
$$	\{{\rm Re}\,\xi_i<c\hbox{ for }i=1,\ldots,m
	\hbox{ and Re}\,\eta_j<c\hbox{ for }j=1,\ldots,n\}      $$
$$	\cup\{{\rm Re}\,\xi_1=\ldots={\rm Re}\,\xi_m<\log r'
	\hbox{ and Re}\,\eta_1=\ldots={\rm Re}\,\eta_n<\log r''\}      $$
for suitable (large negative) $c$.  This is a tube and by the Tube Theorem\footnote{*}{For the standard Tube Theorem see for instance [7] Theorem 2.5.10.  Here we need a variant, the Flattened Tube Theorem, for which see Epstein [1]} $1/\tilde P$ is analytic in
$$	\{{\rm Re}\,\xi_i<\log r'\hbox{ for }i=1,\ldots,m
	\hbox{ and Re}\,\eta_j<\log r''\hbox{ for }j=1,\ldots,n\}      $$
and therefore $\tilde P$ does not vanish when $z_1,\ldots,z_m$ are inside of $\Phi\Gamma'$ and $w_1,\ldots,w_n$ inside $\Phi\Gamma''$.  Going back to the polynomial $P$, we see that it cannot vanish when $\{z_1,\ldots,z_m\}$ and $\{w_1,\ldots,w_n\}$ are separated by $\Gamma'$ and $\Gamma''$.\qed
\medskip
	{\bf 13 Proposition.}
\medskip
	{\sl Suppose that $P(z_1,\ldots,z_n,w_1,\ldots,w_n)$ satisfies the conditions of Proposition A2 and that 
$$	P(z_1,\ldots,z_n,w_1,\ldots,w_n)\ne0      $$
when $|z_1|=\ldots=|z_n|=a$, $|w_1|=\ldots=|w_n|=b$ and $0<a\ne b$.  Then $P$ is a G-nomial.}
\medskip
	Taking $z_1,\ldots,z_n=3/2$, $w_1,\ldots,w_n=1/2$, we have $0\ne P(3/2,\ldots,1/2)=P(1,\ldots,0)$\ $=\alpha$, {\it i.e.}, the coefficient $\alpha$ of the monomial $z_1\ldots z_n$ in $P$ is different from 0 .  Therefore we have
$$	P(z_1,\ldots,z_n,w_1,\ldots,w_n)\ne0\eqno{(6)}      $$
 if $|z_1|=\ldots=|z_n|=a$, $|w_1|=\ldots=|w_n|=b$ and $0\le a<b$;  (6) also holds if $|w_1|=\ldots=|w_n|=b$ provided $|z_1|,\ldots,|z_n|<e^c$ for suitable (large negative) $c$.  Applying the Tube Theorem as in the proof of Proposition 12 we find thus that (6) holds when
$$	|z_1|,\ldots,|z_n|<b\qquad,\qquad|w_1|=\ldots=|w_n|=b      $$
In particular, $P(z_1,\ldots,w_n)\ne0$ if $z_1,\ldots,z_n\in\Gamma'$, $w_1,\ldots,w_n\in\Gamma''$ where $\Gamma'$, $\Gamma''$ are proper circles such that $\Gamma'$ is entirely inside $\Gamma''$ and $\Gamma''$ is centered at 0.  But by conformal invariance (Corollary A3) we can replace these conditions by $\Gamma'\cap\Gamma''=\emptyset$.  Proposition 12 then implies that $P$ is a G-nomial.\qed
\medskip
	{\bf 14 Proposition.}
\medskip
	{\sl Suppose that $P_0(z_1,\ldots,w_n)$ and $P_1(z_1,\ldots,w_n)$ are reduced G-nomials which become equal when $z_1=z_2$:
$$	P_0(z,z,z_3,\ldots,w_n)=P_1(z,z,z_3,\ldots,w_n)      $$
Then, for $0\le\alpha\le1$
$$	P_\alpha(z_1,\ldots,w_n)
	=(1-\alpha)P_0(z_1,\ldots,w_n)+\alpha P_1(z_1,\ldots,w_n)      $$
is again a reduced G-nomial.}
\medskip
	[Note that instead of the pair $(z_1,z_2)$ one could take any pair $(z_i,z_j)$].  
\medskip
	We have to prove that if the proper circle $\Gamma$ separates $\{z_1,\ldots,z_n\}$, $\{w_1,\ldots,w_n\}$, then $P_\alpha(z_1,\ldots,w_n)\ne0$.
\medskip
	Let $p_\alpha(z_1,z_2)$ be obtained from $P_\alpha(z_1,\ldots,w_n)$ by fixing $z_3,\ldots,z_n$ on one side of $\Gamma$ and $w_1,\ldots,w_n$ on the other side.  By assumption
$$	p_\alpha(z_1,z_2)=az_1z_2+b_\alpha z_1+c_\alpha z_2+d\eqno{(7)}      $$
where $b_\alpha=(1-\alpha)b_0+\alpha b_1$, $c_\alpha=(1-\alpha)c_0+\alpha c_1$, and $b_0+c_0=b_1+c_1$.  We have to prove:
\medskip
	(A) {\sl If $z_1,z_2\in\Delta$ where $\Delta$ is the region bounded by $\Gamma$ and not containing $w_1,\ldots,w_n$, then $p_\alpha(z_1,z_2)\ne0$.}
\medskip
	We remark now that, as functions of $z_3,\ldots,w_n$, the expressions
$$	a\qquad,\qquad -{(b_0+c_0)^2\over 4a}+d      $$
cannot vanish identically.  For $a$ this is because the coefficient of $z_1\cdots z_n$ in (7) is 1.  Note now that if we decompose $a$ in prime factors, these cannot occur with an exponent $>1$ because $a$ is of degree $\le1$ in each variable $z_3,\ldots,w_n$.  Therefore if $-(b_0+c_0)^2/4a+d=0$, {\it i.e.}, if $a$ divides $(b_0+c_0)^2$, then $a$ divides $(b_0+c_0)$ and the quotient is homogeneous of degree 1.  But then $(b_0+c_0)^2/4a$ contains some variables with an exponent 2, in contradiction with the fact that in $d$ all variables occur with an exponent $\le1$.  In conclusion $-(b_0+c_0)^2/4a+d$ cannot vanish identically.
\medskip
	By a small change of $z_3,\ldots,w_n$ we can thus assume that 
$$	a\ne0\qquad,\qquad-{(b_0+c_0)^2\over4a}+d\ne0\eqno{(8)}      $$
We shall first consider this case and later use a limit argument to prove (A) when (8) does not hold.  By the change of coordinates
$$	z_1=u_1-{b_0+c_0\over2a}\qquad,\qquad z_2=u_2-{b_0+c_0\over2a}      $$
(linear in $z_1$, $z_2$) we obtain
$$	p_\alpha
=au_1u_2+{1\over2}(b_\alpha-c_\alpha)(u_1-u_2)-{(b_0+c_0)^2\over4a}+d      $$
(Note that $b_\alpha+c_\alpha=b_0+c_0$).  Write
$$	A=(b_0+c_0)^2-4ad\qquad,\qquad\beta={\sqrt A\over2a}
	\qquad,\qquad\lambda(\alpha)={b_\alpha-c_\alpha\over\sqrt A}      $$
for some choice of the square root of $A$, and
$$	u_1=\beta v_1\qquad,\qquad u_2=\beta v_2      $$
then
$$	p_\alpha={A\over4a}(v_1v_2+\lambda(\alpha)(v_1-v_2)-1)      $$
If we write $v_1=(\zeta_1+1)/(\zeta_1-1)$, $v_2=(\zeta_2+1)/(\zeta_2-1)$, the condition $p_\alpha\ne0$ becomes
$$	\zeta_1(1-\lambda(\alpha))
	+\zeta_2(1+\lambda(\alpha))\ne0\eqno{(9)}      $$
\indent
	Note that $\lambda(\alpha)=\pm1$ means $(b_\alpha-c_\alpha)^2-A=0$, {\it i.e.}, $ad-b_\alpha c_\alpha=0$ and
$$	p_\alpha=a(z_1-S_\alpha)(z_2-T_\alpha)      $$
\indent
	By assumption $p_0(z,z)\ne0$ when $z\in\Delta$.  Therefore, the image $\Delta_v$ of $\Delta$ in the $v$-variable does not contain $+1$, $-1$, and the image $\Delta_\zeta$ in the $\zeta$-variable does not contain $0$, $\infty$.  In particular $\Delta_\zeta$ is a circular disc or a half-plane, and thus {\it convex}.
\medskip
	If $\lambda(\alpha)$ is real and $-1\le\lambda(\alpha)\le1$, (9) holds when $\zeta_1,\zeta_2\in\Delta_\zeta$.  [This is because $\Delta_\zeta$ is convex and $\Delta_\zeta\not\ni0$].  Therefore in that case (A) holds.
\medskip
	We may thus exclude the values of $\alpha$ such that $-1\le\lambda(\alpha)\le1$, and reduce the proof of the proposition to the case when at most one of $\lambda(0)$, $\lambda(1)$ is in $[-1,1]$, and the other $\lambda(\alpha)\notin[-1,1]$.  Exchanging possibly $P_0$, $P_1$, we may assume that all $\lambda(\alpha)\notin[-1,1]$ except $\lambda(1)$.  Exchanging possibly $z_1$, $z_2$, ({\it i.e.}, replacing $\lambda$ by $-\lambda$) we may assume that $\lambda(1)\ne1$.  We may finally assume that 
$$	|\lambda(0)+1|+|\lambda(0)-1|
	\ge|\lambda(1)+1|+|\lambda(1)-1|\eqno{(10)}      $$
where the left hand side is $>2$ while the right hand side is =2 if $\lambda(1)\in[-1,1]$.
\medskip
	For $\alpha\in[0,1]$ we define the map
$$	f_\alpha:
	\zeta\mapsto{\lambda(\alpha)+1\over\lambda(\alpha)-1}\zeta      $$
When $\alpha=0,1$ the inequality (9) holds by assumption for $\zeta_1,\zeta_2\in\Delta_\zeta$.  [Note that the point $v=\infty$, {\it i.e.}, $\zeta=1$ does not make a problem: if $\lambda\ne\pm1$ this is seen by continuity; if $\lambda=\pm1$ this follows from $\Delta_\zeta\not\ni0$].  Therefore
$$	\Delta_\zeta\cap f_0\Delta_\zeta=\emptyset\qquad,
	\qquad\Delta_\zeta\cap f_1\Delta_\zeta=\emptyset      $$
We want to show that $\Delta_\zeta\cap f_\alpha\Delta_\zeta=\emptyset$ for $0<\alpha<1$.  In fact, it suffices to prove 
$$	\Delta'_\zeta\cap f_\alpha\Delta'_\zeta=\emptyset      $$
for slightly smaller $\Delta'\subset \Delta_\zeta$, {\it viz}, the inside of a proper circle $\Gamma'$ such that 0 is outside of $\Gamma'$.  Since we may replace $\Delta'$ by any $c\Delta'$ where $c\in{\bf C}\backslash\{0\}$, we assume that $\Delta'$ is the interior of a circle centered at $\lambda(0)-1$ and with radius $r^0_-<|\lambda(0)-1|$.  Then $f_0\Delta'$ is the interior of a circle centered at $\lambda(0)+1$ and with radius $r^0_+$.  The above two circles are disjoint, but we may increase $r^0_-$ until they touch, obtaining
$$	r^0_-+r^0_+=2\qquad,\qquad
	 r^0_+=|{\lambda(0)+1\over\lambda(0)-1}|r^0_-      $$
{\it i.e.},
$$	r_-^0={2|\lambda(0)-1|\over|\lambda(0)+1|+|\lambda(0)-1|}\qquad,
	\qquad r_+^0={2|\lambda(0)+1|\over|\lambda(0)+1|+|\lambda(0)-1|}     $$
We define $r_-^\alpha$ and $r_+^\alpha$ similarly, with $\lambda(0)$ replaced by $\lambda(\alpha)$ for $\alpha\in[0,1]$.  To prove that $\Delta'\cap f_\alpha\Delta'=\emptyset$ for $0<\alpha<1$, we may replace $\Delta'$ by ${\lambda(\alpha)-1\over\lambda(0)-1}\Delta'$ (which is a disc centered at $\lambda(\alpha)-1$) and it suffices to prove that the radius $|{\lambda(\alpha)-1\over\lambda(0)-1}|r_-^0$ of this disc is $\le r_-^\alpha$, {\it i.e.},
$$	{2|\lambda(\alpha)-1|\over|\lambda(0)+1|+|\lambda(0)-1|}
\le{2|\lambda(\alpha)-1|\over|\lambda(\alpha)+1|+|\lambda(\alpha)-1|}      $$
or
$$	|\lambda(0)+1|+|\lambda(0)-1|
	\ge|\lambda(\alpha)+1|+|\lambda(\alpha)-1|\eqno{(11)}      $$
Note now that $\{\lambda\in{\bf C}:|\lambda+1|+|\lambda-1|=const.\}$ is an ellipse with foci $\pm1$, and since $\lambda(\alpha)$ is affine in $\alpha$, the maximum value of $|\lambda(\alpha)+1|+|\lambda(\alpha)-1|$ for $\alpha\in[0,1]$ is reached at 0 or 1, and in fact at 0 by (10).  This proves (11).
\medskip
	This concludes the proof of (A) under the assumption (8).  Consider now a limiting case when (8) fails and suppose that (A) does not hold.  Then, by Lemma A1, $p_\alpha$ vanishes identically.  In particular this would imply $p_\alpha(z,z)=0$, in contradiction with the assumption that $P_0$ is a G-nomial.
\medskip
	We have thus shown that $P_\alpha$ is a G-nomial, and since it is homogeneous of degree $n$ in the $2n$ variables $z_1,\ldots,w_n$, and contains $z_1\cdots z_n$ with coefficient 1, $P_\alpha$ is a reduced G-nomial.\qed
\medskip
	{\bf 15 Corollary} (contractibility).
\medskip
	{\sl The set $G_n$ of reduced G-nomials is contractible.}
\medskip
	In the linear space of MA-nomials $P(z_1,\ldots,w_n)$ satisfying the conditions of Proposition A2 we define a flow by
$$	{dP\over dt}
=-P+\textstyle{({n\atop 2})}^{-1}\sum\textstyle{^*}\pi P\eqno{(12)}      $$
where $\Sigma^*$ is the sum over the $({n\atop 2})$ transpositions $\pi$, {\it i.e.} interchanges of two of the variables $z_1,\ldots,z_n$ of $P$.  In view of Proposition 14, the positive semiflow defined by (12) preserves the set $G_n$ of reduced G-nomials.  Condition (b)$_n$ of Proposition A2 shows that the only fixed point of (12) is, up to a normalizing factor, Grace's polynomial $P_\Sigma$.  We have thus a contraction of $G_n$ to $\{P_\Sigma\}$, and $G_n$ is therefore contractible.\qed
\bigskip
\centerline{\bf A. Appendix.}
\bigskip
	{\bf A1 Lemma} (limits).
\medskip
	{\sl Let $D_1,\ldots,D_r$ be open discs, and assume that the MA-nomials $P_k(z_1,\ldots,z_r)$ do not vanish when $z_1\in D_1$, \dots, $z_r\in D_r$.  If the $P_k$ have a limit $P_\infty$ when $k\to\infty$, and if $P_\infty(\hat z_1,\ldots,\hat z_r)=0$ for some $\hat z_1\in D_1$, \dots, $\hat z_r\in D_r$, then $P_\infty=0$ identically.}
\medskip
	There is no loss of generality in assuming that $\hat z_1=\ldots=\hat z_r=0$.  We prove the lemma by induction on $r$.  For $r=1$, if the affine function $P_\infty$ vanishes at 0 but not identically, the implicit function theorem shows that $P_k$ vanishes for large $k$ at some point close to 0, contrary to assumption.  For $r>1$, the induction assumption implies that putting any one of the variables $z_1,\ldots,z_r$ equal to 0 in $P_\infty$ gives the zero polynomial.  Therefore $P_\infty(z_1,\ldots,z_r)=\alpha z_1\cdots z_r$.  Fix now $z_j=a_i\in D_i\backslash\{0\}$ for $j=1,\ldots,r-1$.  Then $P_k(a_1,\ldots,a_{r-1},z_r)\ne0$ for $z_r\in D_r$, but the limit $P_\infty(a_1,\ldots,a_{r-1},z_r)=\alpha a_1\cdots a_{r-1}z_r$ vanishes at $z_r=0$ and therefore identically, {\it i.e.}, $\alpha=0$, which proves the lemma.\qed
\medskip

	{\bf A2 Proposition} (reduced forms).
\medskip
	{\sl For $n\ge1$, the following conditions on a MA-nomial $P(z_1,\ldots,z_n,w_1,\ldots,w_n)$ not identically zero are equivalent:
\medskip
	(a)$_n$ $P$ satisfies}
$$	P(z_1+\xi,\ldots,w_n+\xi)=P(z_1,\ldots,w_n)\qquad
	\qquad\hbox{(translation\quad invariance)}      $$
$$	P(\lambda z_1,\ldots,\lambda w_n)=\lambda^nP(z_1,\ldots,w_n)\qquad
	\qquad\hbox{(homogeneity of degree $n$)}      $$
\indent{\sl
	(b)$_n$ There are constants $C_\pi$ such that 
$$	P(z_1,\ldots,w_n)=\sum_\pi C_\pi\prod_{j=1}^n(z_j-w_{\pi(j)})      $$
where the sum is over all permutations $\pi$ of $(1,\ldots,n)$.}
\medskip
	We say that (b)$_n$ gives a {\it reduced form} of $P$ (it need not be unique).
\medskip
	Clearly (b)$_n\Rightarrow$(a)$_n$.  We shall prove (a)$_n\Rightarrow$(b)$_n$ by induction on $n$, and obtain at the same time a bound $\sum|C_\pi|\le k_n.||P||$ for some norm $||P||$ (the space of $P$'s is finite dimensional, so all norms are equivalent).  Clearly, (a)$_1$ implies that $P(z_1,w_1)=C(z_1-w_1)$, so that (b)$_1$ holds.  Let us now assume that $P$ satisfies (a)$_n$ for some $n>1$.
\medskip
	If $X$ is an $n$-element subset of $\{z_1,\ldots,w_n\}$, let $A(X)$ denote the coefficient of the corresponding monomial in $P$.  We have
$$	\sum_XA(X)=P(1,\ldots,1)=P(0,\ldots,0)=0      $$
In particular 
$$	\max_{X',X''}|A(X')-A(X'')|\ge\max_XA(X)      $$
Note also that one can go from $X'$ to $X''$ in a bounded number of steps exchanging a $z_j$ and a $w_k$.  Therefore one can choose $z_j$, $w_k$, $Z$ containing $z_j$ and not $w_k$, and $W$ obtained by replacing $z_j$ by $w_k$ in $Z$ so that 
$$	|A(Z)-A(W)|\ge\alpha(\sum_X|A(X)|^2)^{1/2}      $$
where $\alpha$ depends only on $n$.
\medskip
	Write now
$$	P=az_jw_k+bz_j+cw_k+d      $$
where the polynomials $a$, $b$, $c$, $d$ do not contain $z_j$, $w_k$.  We have thus
$$	P=P_1+{1\over2}(b-c)(z_j-w_k)      $$
where 
$$	P_1=az_jw_k+{1\over2}(b+c)(z_j+w_k)+d      $$
Let $\tilde a$, $\tilde b$, $\tilde c$, $\tilde d$ be obtained by adding $\xi$ to all the arguments of $a$, $b$, $c$, $d$.  By translation invariance we have thus
$$	az_jw_k+bz_j+cw_k+d
=\tilde a(z_j+\xi)(w_k+\xi)+\tilde b(z_j+\xi)+\tilde c(w_k+\xi)+\tilde d  $$
$$	=\tilde az_jw_k+(\tilde a\xi+\tilde b)z_j+(\tilde a\xi+\tilde c)w_k
	+\tilde a\xi^2+(\tilde b+\tilde c)\xi+\tilde d      $$
hence $\tilde b-\tilde c=b-c$.  Therefore $b-c$ satisfies (a)$_{n-1}$ and, using the induction assumption we see that
$$	{1\over2}(b-c)(z_j-w_k)      $$
has the form given by (b)$_n$.  In particular $P_1$ again satisfies (a)$_n$.
\medskip
	We compare now the coefficients $A_1(X)$ for $P_1$ and $A(X)$ for $P$:
$$	\sum_X|A(X)|^2-\sum_X|A_1(X)|^2
	\ge|A(Z)|^2+|A(W)|^2-{1\over2}|A(Z)+A(W)|^2      $$
$$	={1\over2}|A(Z)-A(W)|^2\ge{\alpha^2\over2}\sum_X|A(X)|^2      $$
so that
$$	\sum|A_1(X)|^2\le(1-{\alpha^2\over2})\sum_X|A(X)|^2      $$
We have thus a geometrically convergent approximation of $P$ by expressions satisfying (b)$_n$, and an estimate of $\sum|C_\pi|$ as desired.\qed
\medskip
	{\bf A3 Corollary}
\medskip
	{\sl If the MA-nomial $P(z_1,\ldots,w_n)$ satisfies the conditions of Proposition A2, the following properties hold:
\medskip
	{\rm (conformal invariance)} if $ad-bc\ne0$, then
$$	P({az_1+b\over cz_1+d},\ldots,{aw_n+b\over cw_n+d})
	=P(z_1,\ldots,w_n)\prod_{j=1}^n{ad-bc\over(cz_j+d)(cw_j+d)}      $$
\indent
	{\rm (roots)} the polynomial
$$	\hat P(z)=P(z,\ldots,z,w_1,\ldots,w_n)      $$
has exactly the roots $w_1,\ldots,w_n$ (repeated according to multiplicity).}
\medskip
	These properties follow directly if one writes $P$ in reduced form.\qed
\medskip
	{\bf References.}

[1] H. Epstein.  ``Some analytic properties of scattering amplitudes in quantum field theory.'' in M. Chretien and S. Deser eds. {\it Axiomatic Field Theory.}  Gordon and Breach, New York, 1966.

[2] O.J. Heilmann and E.H. Lieb.  ``Theory of monomer-dimer systems.'' 
Commun. Math. Phys. {\bf 25},190-232(1972); {\bf 27},166(1972).

[3] T. D. Lee and C. N. Yang.  ``Statistical theory of equations of state
and phase relations. II.  Lattice gas and Ising model.''  Phys. Rev. {\bf
87},410-419(1952).

[4] G. Polya and G. Szeg\"o.  {\it Problems and theorems in analysis II.}  Springer, Berlin, 1976.

[5] D. Ruelle.  ``Extension of the Lee-Yang circle theorem.''  Phys. Rev.
Letters {\bf 26},303-304(1971).

[6] D. Ruelle.  ``Counting unbranched subgraphs.''  J. Algebraic Combinatorics {\bf 9},157-160(1999);  ``Zeros of graph-counting polynomials.''  Commun. Math. Phys. {\bf 200},43-56(1999). 

[7] L. H\"ormander.  {\it An introduction to complex analysis in several variables.}  D. Van Nostrand, Princeton, 1966.

\end